\documentclass[12pt,a4paper]{article}
\usepackage{cite}
\usepackage{axodraw}
\let\vec\mathbf                                   
\def\Dirac{\not}                                  
\def\dirac{\not\!}                                
\def\vprod{\mathbin{\null_\wedge}}                
\def\tr{\mathrm{tr}}                              
\def\QED{\mathrm{QED}}                            
\def\QCD{\mathrm{QCD}}                            
\def\Order{\mathrm{O}}
\def\Odderon{\mathrm{O}}
\def\Pomeron{\mathrm{P}}

\title{
  \hfill\raisebox{2cm}[0pt][0pt]{\normalsize EPTCO-99-001} \\ 
  Single-Spin Asymmetries \\
  for Small-Angle Pion Production \\
  in High-Energy Hadron Collisions \\
  {\small (submitted to \emph{Eur.\ Phys.\ J} \textbf{C})}}

\author{
  Azad Ahmedov$^{1,}$\thanks{E-mail: \texttt{ahmedov@sunse.jinr.ru}}\space,
  Igor V. Akushevich$^{2,}$\thanks{E-mail: \texttt{aku@hep.by}}\space,
\\
  Eduard A. Kuraev$^{1,}$\thanks{E-mail: \texttt{kuraev@thsun1.jinr.ru}}\space\,
  and
  Philip G. Ratcliffe$^{3,}$\thanks{E-mail: \texttt{pgr@fis.unico.it}}
\\[12pt]
  \hskip-1.5em
  \parbox{\textwidth}
  {\begin{list}{$^\arabic{enumi}$}
               {\small
                \topsep0pt
                \partopsep0pt
                \parsep0pt
                \itemsep0pt
                \usecounter{enumi}%
                \labelwidth1.1ex
                \def\makelabel##1{##1\hss}%
                \it
               }
  \item Joint Inst.\ for Nuclear Research,
        141980, Dubna, Moscow region, Russia
  \item Nat.\ Center of Particle and High Energy Physics,
        Bogdanovich Str.~153, 220040 Minsk, Belarus
  \item Dip.\ di Scienze CC.FF.MM., Univ.\ degli Studi dell'Insubria,
        via Lucini 3, 22100 Como, Italy
        and Ist.\ Naz.\ di Fisica Nucleare---sezione di Milano
  \end{list}}}

\date{February 1999}
\begin{document}
\maketitle
\begin{abstract}

  Within the framework of a simple model, we study single-spin
  asymmetries for pion production in hadron-hadron collisions at
  high-energies with one hadron polarised. The asymmetries are
  generated via a mechanism of final (initial) state interactions. For
  peripheral kinematics, when the pion belongs to the fragmentation
  region of the polarised proton, we find non-zero asymmetries in the
  high-energy limit. Numerical results and comparision with existing
  experimental data are presented. We also discuss the relationship
  with odderon exchange phenomenology.

\end{abstract}
\newpage
\section{Introduction}

Single-spin correlations have been the subject of theoretical
\cite{Kane:1978nd, Qiu:1991pp, Qiu:1992wg, Barni:1992qn, Ratcliffe:1992rh,
  Arbuzov:1994x1, 
  Efremov:1995dg, Korotkiian:1995vf, Anselmino:1995tv, Anselmino:1996x4,
  Hammon:1996pw, Qiu:1999ia, Anselmino:1998yz, Hammon:1998gb,
  Ratcliffe:1998su} and experimental \cite{Adams:1991cs, Adams:1991rw,
  Adams:1992ru, Bravar:1995fw, Adams:1997wf} study since the
seventies. Earlier theoretical work paid attention mainly to
time-reversal invariance violation in hadron-scattering processes. No
such effect was found whereas C-odd single-spin correlation
asymmetries had been observed at the level of ${\sim}10\%$ in SLAC
experiments with 10--12\,GeV electrons and positrons scattering off
polarised protons, with, however, large error bars and later in
Fermilab experiments at higher energies. The capabilities of modern
CERN and DESY experiments permit the reduction of these errors due to
much improved statistics. We argue here that, at small momentum
transfer, large effects may be understood in the framework of pomeron
and odderon exchange models and may thus provide an independent method
of studying the characteristics of such exchanges in high-energy
pheripheral hadron scattering.

The appearance of single-spin correlations and associated asymmetries 
in differential
cross-sections is due to a quantum effect of interference between 
real and imaginary parts of different amplitudes. In phenomenological
approaches, the amplitudes have been used in a Breit-Wigner form and
the asymmetries turn out to be proportional to the width-to-mass ratio
of the resonance \cite{Barni:1992qn, Ratcliffe:1992rh}. For the case
of polarised proton-proton collisions, with the production of pions
through some intermediate nucleon resonance state in peripheral
kinematics (PK), the asymmetry may be as large as 20--40\%
\cite{Arbuzov:1994x1}. Another mechanism for the generation of
imaginary parts in scattering amplitudes is due to initial- or
final-state interactions. In lowest-order perturbation theory such
contributions can arise from the interference between the Born
amplitude and one-loop amplitudes with a non-zero s-channel imaginary
part \cite{Kane:1978nd}.

A similar phenomena has been found for the case of large-angle
production: for the kinematics of large $p_T$ and large $x_F$ of the
detected hadron, single-spin asymmetries are generated by twist-3
parton correlation functions constructed from quark and gluon fields.
For this case, the theoretical approach within the framework of
perturbative QCD has been recently discussed by several authors
\cite{Qiu:1991pp, Qiu:1992wg, Efremov:1995dg, Qiu:1999ia,
  Ratcliffe:1998su}.

Here we consider peripheral kinematics, when the unpolarised proton
produces a jet moving along the initial direction of motion, which is
then not detected, whereas the jet produced by the polarised proton
contains a detected pion. The asymmetry originates from a term
$i\epsilon^{p_1p_2la}=\frac{i}{2}s[\vec{l}\vprod\vec{a}]_z$, where
$p_1$ and $p_2$ are the 4-momenta of the initial protons, $l$ is the
momentum of the pion in the center-of mass system (CMS) and $a$ is the
spin 4-vector of the proton with momentum $p_2$, which is essentially
a 2-component vector located in the plane transverse to the beam axis
(z-direction of the initial-state proton with momentum $p_1$), and
$s=4E^2$ is the square of the total CMS energy.

For high enough energies, the description in terms of Regge
trajectories is more convenient since, for small enough momentum
transfer, the contribution of operators of all twists will be of the
same order of magnitude. Here the even (spin-independent) part of the
differential cross-section is determined by Pomeron exchange, whereas
the spin-dependent part, arising from interference between one- and
two-gluon exchange amplitudes, should be described by the odderon
trajectory.  Note that quark exchange in the $t$-channel only gives a
small contribution, suppressed by a factor of $m^2/s$, where $m$ is
the proton mass. Thus, the study of single-spin asymmetries in PK may
provide information on the odderon intercept. In this paper, using a
QED-like framework, with the point-like hadrons, we calculate the
asymmetry, defined as follows:
\begin{eqnarray}
  A
  &=&
  \frac{d\sigma(a,l)-d\sigma(-a,l)}
       {d\sigma(a,l)+d\sigma(-a,l)}
\\
  &=&
  \alpha_\QED \frac{(\vec{a}\vprod\vec{l})_z}{m} \mathcal{A}(r,x),
\end{eqnarray}
where $\alpha_\QED=1/137$ is the QED coupling constant and the
resolving power, $\mathcal{A}(r,x)$, is a function of $r=l_\bot/m$
(the transverse momentum of pion in units of the proton mass, $m$) and
$x=2l_0/\sqrt{s}$ (its energy fraction, $x\sim\Order(1)$).  We shall
show that $\mathcal{A}(r,x)$ is a smooth rising function of $x$ of
order unity. Thus, naively replacing the QED coupling constant by that
of QCD or by the pomeron or odderon coupling with the proton, we find
that the asymmetry may be large for large enough values of
$l_\bot$, in qualitative agreement with experimental data
\cite{Adams:1991cs, Adams:1992ru, Adams:1991rw, Adams:1997wf}.

The paper is organised as follows. In Section~\ref{sec:CalcCS} we
calculate the expressions for the squared matrix elements summed over
spin states for processes of neutral- and charged-pion production in
the framework of our QED-like approach. The
corresponding charge-odd interferences for these channels are
considered in Section~\ref{sec:CalcSD}, where first we obtain the
expressions for asymmetries in an exclusive set-up (when both the nucleon
and pion from the jet developing along the direction of polarised
proton are fixed in the experiment) and then the ratio of odd and even
parts of the cross-section averaged over final proton momenta are
estimated. In conclusion, we discuss the situation when a hadron is
detected in the opposite direction and also discuss the role of
higher-order perturbation theory contributions and the relationship to
odderon exchange.

\section{Calculation of the Cross-Section}
\label{sec:CalcCS}

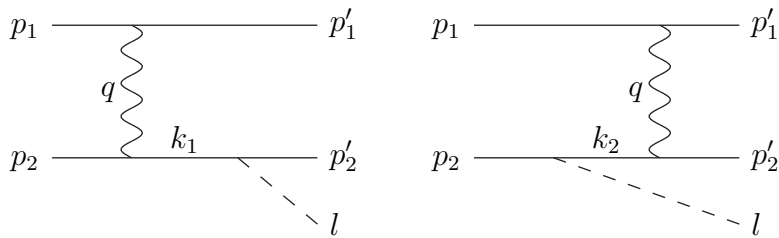
\begin{figure}[htbp]
  \begin{center}
\begin{picture}
          (140, 75)(-20,  0)
\Line     (  0, 75)(100, 75)
\Line     (  0, 25)(100, 25)
\Photon   ( 30, 75)( 30, 25) {4} {4}
\DashLine ( 70, 25)(100,  0) {5}
\Text     ( -5, 70)[br]{$p_1$}
\Text     ( -5, 20)[br]{$p_2$}
\Text     (105, 70)[bl]{$p_1'$}
\Text     (105, 20)[bl]{$p_2'$}
\Text     (105,  0)[ l]{$l$}
\Text     ( 24, 50)[ r]{$q$}
\Text     ( 50, 27)[b ]{$k_1$}
\end{picture}
\hspace{1em}
\begin{picture}
          (140, 75)(-20,  0)
\Line     (  0, 75)(100, 75)
\Line     (  0, 25)(100, 25)
\Photon   ( 70, 75)( 70, 25) {4} {4}
\DashLine ( 30, 25)(100,  0) {5}
\Text     ( -5, 70)[br]{$p_1$}
\Text     ( -5, 20)[br]{$p_2$}
\Text     (105, 70)[bl]{$p_1'$}
\Text     (105, 20)[bl]{$p_2'$}
\Text     (105,  0)[ l]{$l$}
\Text     ( 64, 50)[ r]{$q$}
\Text     ( 50, 27)[b ]{$k_2$}
\end{picture}
 
    \caption{\label{fig:single}
      The amplitudes for the process $pp\to{}pp\pi$ (a) in the Born
      approximation; the proton with momentum $p_2$ is polarised.}
  \end{center}
\end{figure}

Consider first the process
\begin{equation}
  P(p_1) + P(p_2) \to P(p_1') + P(p_2') + \pi_0(l),
\end{equation}
for which the relevant Born-approximation diagrams are show in
fig.~\ref{fig:single}.
We shall use the Sudakov expansion for the momenta of the problem,
introducing the almost light-like vectors
\begin{eqnarray}
  \tilde{p}_1^{\mu}
  &=&
  p_1^{\mu} - \frac{m^2}{s}p_2^{\mu},
\nonumber\\
  \tilde{p}_2^{\mu}
  &=&
  p_2^{\mu} - \frac{m^2}{s}p_1^{\mu}.
\end{eqnarray}

With these we define the following parametrisation of the momenta in
the problem:
\begin{eqnarray}
  q^{\mu}
  &=&
  p_1^{\mu}-p_1^{'\mu}
  \;\,=\;\,
   \alpha \tilde{p}_2^{\mu}
  + \beta \tilde{p}_1^{\mu}
  +            q_\bot^{\mu},
\nonumber\\
  p_2^{'\mu}
  &=&
  (1-x) \tilde{p}_2^{\mu} + \beta' \tilde{p}_1^{\mu} + p_\bot^{\mu},
\nonumber\\
  l^{\mu}
  &=&
  x \tilde{p}_2^{\mu} + \beta_l \tilde{p}_1^{\mu} + l_\bot^{\mu},
\nonumber\\
  q_\bot^{\mu}
  &=&
  p_\bot^{\mu} + l_\bot^{\mu},
\end{eqnarray}
The transverse parts, $v_\bot$, obey
\begin{eqnarray}
  v_\bot{\cdot}p_1
  &=&
  0 \;\,=\;\, v_\bot{\cdot}p_2,
\nonumber\\
  v_\bot^2
  &=&
  -\vec{v}^2 \;\,<\;\, 0.
\end{eqnarray}
We have also used the specific properties of PK: the sum of energy
fractions of pion and proton from the jet moving along initial
polarised proton is equal to unity, $x\sim\Order(1)$; moreover, the
reality conditions for final particles permits the neglect of the
``small'' components of momenta:
\begin{eqnarray}
  s\beta'
  &=&
  \frac{m^2+\vec{p}_2^2}{1-x},
\\
  s\beta_l
  &=&
  \frac{\vec{l}^2}{x}.
\end{eqnarray}

Here and in what follows we neglect the pion mass squared compared to
that of the proton. The intermediate fermion denominators in the Born
graphs (see Fig.~\ref{fig:single}) are then given by
\begin{eqnarray}
  d_1
  &=&
  k_1^2 - m^2
  \;\,\simeq\;\, \hphantom{-}
  \frac{m^2x^2+(x\vec{q}-\vec{l})^2}
       {x(1-x)},
\\
  d_2
  &=&
  k_2^2 - m^2
  \;\,\simeq\;\,
  - \frac{m^2x^2+\vec{l}^2}{x}.
\end{eqnarray}
Note that $k_1^2$ is just the invariant-mass squared of the jet moving
along $\vec{p}_2$.  We shall show that the dominant contribution
arises when this quantity is of the order of some nucleon mass
squared.

In the Born approximation, the matrix element has the following form:
\begin{equation}
  \mathcal{M}_{\pi^0}
  \;\, = \;\,
  \frac{4\pi\alpha g}{q^2}
  J^{(1)}_\mu(p_1) D^{\mu\nu}(q) J^{(2)}_\nu(p_2),
\end{equation}
where $\alpha=\alpha_\QED$, $g$ is the pion-nucleon coupling constant
and $D^{\mu\nu}$ is the exchange photon polarisation tensor. The
current vectors introduced are
\begin{eqnarray}
  J^{(1)}_\mu(p_1)
  &=&
  \bar{u}(p_1')\gamma_{\mu} u(p_1),
\nonumber\\
  J^{(2)}_\nu(p_2)
  &=&
  \bar{u}(p_2')O_{\nu} u(p_2),
\end{eqnarray}
where
\begin{eqnarray}
  O_\nu
  &=&
  \frac{1}{d_1} \gamma_5   (\dirac{p}_2'+\Dirac{l}+m) \gamma_\nu
 +\frac{1}{d_2} \gamma_\nu (\dirac{p}_2 -\Dirac{l}+m) \gamma_5
\nonumber\\
  &=&
  \gamma_5
  \left[ \frac{\Dirac{l}\gamma_\nu}{d_1}
       - \frac{\gamma_\nu\Dirac{l}}{d_2} \right].
\label{eq:Onu}
\end{eqnarray}
In the last step of (\ref{eq:Onu}) we have used the Dirac equation for
free protons.  For PK, only the so-called ``nonsense'' components of
the decomposition of photon polarisation tensor give a non-vanishing
contribution in the high-energy limit:
\begin{eqnarray}
  D^{\mu\nu}
  &=&
  g_\bot^{\mu\nu}
  + 2(p_1^\mu p_2^\nu
    + p_1^\nu p_2^\mu)/s
\nonumber\\
  &\simeq&
  2 p_1^\nu p_2^\mu/s.
\end{eqnarray}
Further simplifications may be made using the current conservation
condition, $q{\cdot}J^{(2)}=0$, which implies
$p_1{\cdot}J^{(2)}\simeq-q_\bot{\cdot}J^{(2)}/\beta$.

As a result, the matrix element in the Born approximation becomes
\begin{equation}
  \mathcal{M}_{\pi^0}
  \;\, = \;\,
  -\frac{8\pi\alpha g}{\beta s q^2} \;
   \bar{u}(p_1') \dirac{p}_2 u(p_1) \;
   \bar{u}(p_2') O u(p_2),
\end{equation}
where $O=q_\bot^\nu{}O_\nu$. Here the exchange denominator is
$q^2=-(\vec{q}^2+q_m^2)$, with $q_m^2=m^2(\tilde{s}/s)^2$ and
$\tilde{s}=(m^2x+\vec{p}_2^2)/(1-x)+\vec{l}^2/x$. However, in order to
regulate the infrared divergences, we shall use a massive
vector-particle propagator: $q^2=-(\vec{q}^2+\mu^2)$.  For the modulus
squared of the Born matrix element, summed over spin states, we thus
obtain
\begin{equation}
  \sum_\mathrm{spins}|\mathcal{M}_{\pi_0}|^2
  \;\, = \;\, (16\pi g\alpha)^2
  \frac{s^2 x^2 \vec{q}^2}
       {d_1 (-d_2) (\vec{q}^2 + \mu^2)^2}.
\end{equation}
Note that here we consider the hadrons as point-like particles and so 
the high-frequency contributions are not negligible (unrealistically so). 
Thus, although all integrals are in fact convergent, we introduce a 
form-factor of the form $\exp(-b\vec{q}^2)$, with $b\sim1\,$GeV$^{-2}$.
As is well known, single-spin correlation effects are
absent in the Born approximation, as a consequence of the reality of
Born amplitudes and the form of the proton spin-density matrix:
\begin{eqnarray}
  u(p_2,a)\bar{u}(p_2,a)
  &=&
  (\dirac{p}_2 + m) (1 + \gamma_5\dirac{a}),
\nonumber\\
  \tr[\gamma_5\dirac{a}\dirac{b}\dirac{c}\dirac{d}]
  &=&
  4i \epsilon^{abcd}.
\end{eqnarray}

It is also well known that in the case of elastic small-angle
charged-particle scattering, the Born amplitude acquires a Coulomb
phase factor, $\exp[i\alpha\pi\ln(-q^2/\mu^2)]$, when multiphoton
exchange is taken into account. A similar factor appears in the case
of inelastic processes in PK, such as those we are considering here.

\section{Calculation of the Spin Dependence}
\label{sec:CalcSD}

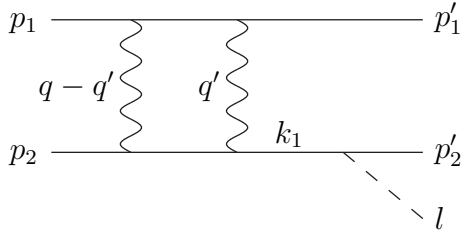
\begin{figure}[htbp]
  \begin{center}
\begin{picture}
          (180, 75)(-20,  0)
\Line     (  0, 75)(140, 75)
\Line     (  0, 25)(140, 25)
\Photon   ( 30, 75)( 30, 25) {4} {4}
\Photon   ( 70, 75)( 70, 25) {4} {4}
\DashLine (110, 25)(140,  0) {5}
\Text     ( -5, 70)[br]{$p_1$}
\Text     ( -5, 20)[br]{$p_2$}
\Text     (145, 70)[bl]{$p_1'$}
\Text     (145, 20)[bl]{$p_2'$}
\Text     (145,  0)[ l]{$l$}
\Text     ( 24, 50)[ r]{$q-q'$}
\Text     ( 64, 50)[ r]{$q'$}
\Text     ( 90, 27)[b ]{$k_1$}
\end{picture}
 
    \caption{\label{fig:double}
      An example higher-order contribution to the process
      $pp\to{}pp\pi$, as considered in the text.}
  \end{center}
\end{figure}

For the spin-dependent part of the interference between single- and
double-photon exchange amplitudes (see Fig.~\ref{fig:double}),
a calculation similar to that performed above gives
\begin{eqnarray}
  \sum_\mathrm{spins}\Delta|\mathcal{M}_{\pi_0}|^2
  &=&
  \frac{ 2^{11} \pi^2 \alpha^3 g^2 s}{\tilde{s}^2 |\vec{q}|^4}
  \ln\left(\frac{|\vec{q}|^2}{\mu^2}\right)
\nonumber\\
  &\times&
  \frac{i}{4} \tr[-\gamma_5 \dirac{a} \tilde{O} (\dirac{p}_2'+m)
                  O(\dirac{p}_2'+m)\dirac{p}_1(\dirac{p}_2+m)],
\end{eqnarray}
where again $O=q_\bot^\nu{}O_\nu$. After calculating the trace we
obtain, for the exclusive set-up,
\begin{eqnarray}
  A
  &=&
  \frac{\Delta\sum|\mathcal{M}_{\pi_0}|^2}
       {      \sum|\mathcal{M}_{\pi_0}|^2} \nonumber\\
  &=&
  4\alpha\ln(\vec{q}^2/\mu^2) m |\vec{a}| |\vec{q}|\sin\phi_q
  \frac{[ x\tilde{s} - 2\vec{q}\cdot\vec{l} ]
        [ x\vec{q}^2 + (1-x)2\vec{q}\cdot\vec{l} ]}
       {\tilde{s}\,d_2\,\vec{q}^2}, \qquad
\label{eq:asym}
\end{eqnarray}
where $\phi_q$ is the azimuthal angle between the transverse 2-vectors
$\vec{a}$ and $\vec{q}$.  We note that the asymmetry is finite in the
small-$\vec{q}$ limit. 

\begin{figure}[htbp]
  \unitlength 1mm
  \begin{center}
\begin{picture}(160,70)
\put(35,0){\makebox(0,0){(a)}}
\put(100,0){\makebox(0,0){(b)}}
\put(90,56.5){\makebox(0,0){\scriptsize (3) -- $x$=0.65}}
\put(90,53){\makebox(0,0){\scriptsize (2) -- $x$=0.40}}
\put(90,49.5){\makebox(0,0){\scriptsize (1) -- $x$=0.20}}
\put(62,40){\makebox(0,0){\scriptsize (3)}}
\put(62,36){\makebox(0,0){\scriptsize (2)}}
\put(62,31){\makebox(0,0){\scriptsize (1)}}
\put(25,45){\makebox(0,0){\scriptsize (3) -- $q_c$=1.4 GeV}}
\put(25,41){\makebox(0,0){\scriptsize (2) -- $q_c$=1.2 GeV}}
\put(25,37){\makebox(0,0){\scriptsize (1) -- $q_c$=1.0 GeV}}
\put(85,13.5){\makebox(0,0){\scriptsize (1)}}
\put(85,26){\makebox(0,0){\scriptsize (2)}}
\put(87,37.5){\makebox(0,0){\scriptsize (3)}}
\put(115,55){\makebox(0,0){\scriptsize $p+p\rightarrow \pi^0+X$}}
\put(123,51){\makebox(0,0){\scriptsize \cite{Adams:1991cs}}}
\put(30,54.5){\makebox(0,0){\scriptsize $p+p\rightarrow \pi^++X$
\cite{Adams:1992ru}}}
\put(30,50){\makebox(0,0){\scriptsize $p+p\rightarrow \pi^0+X$
\cite{Adams:1991cs}}}
\put(60,4){\makebox(0,0){\small $x$}}
\put(120,4){\makebox(0,0){\small $l_\bot$ (GeV)}}
\put(15,63){\makebox(0,0){\small $A$ (\%)}}
\put(80,63){\makebox(0,0){\small $A$ (\%)}}
\put(0,0){
  \epsfxsize=7cm
  \epsfysize=7cm
  \epsfbox{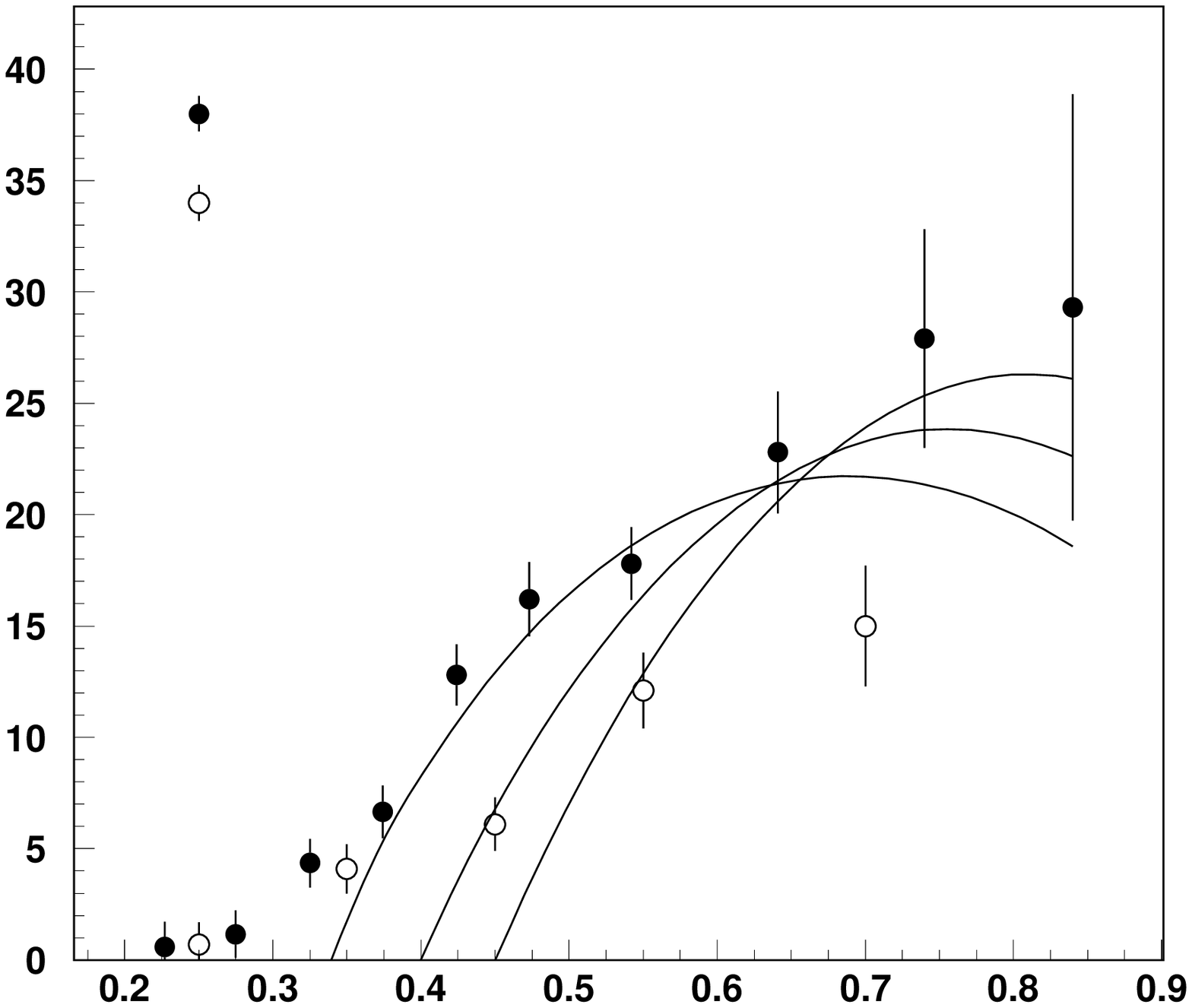}
  }
\put(65,0){
  \epsfxsize=7cm
  \epsfysize=7cm
  \epsfbox{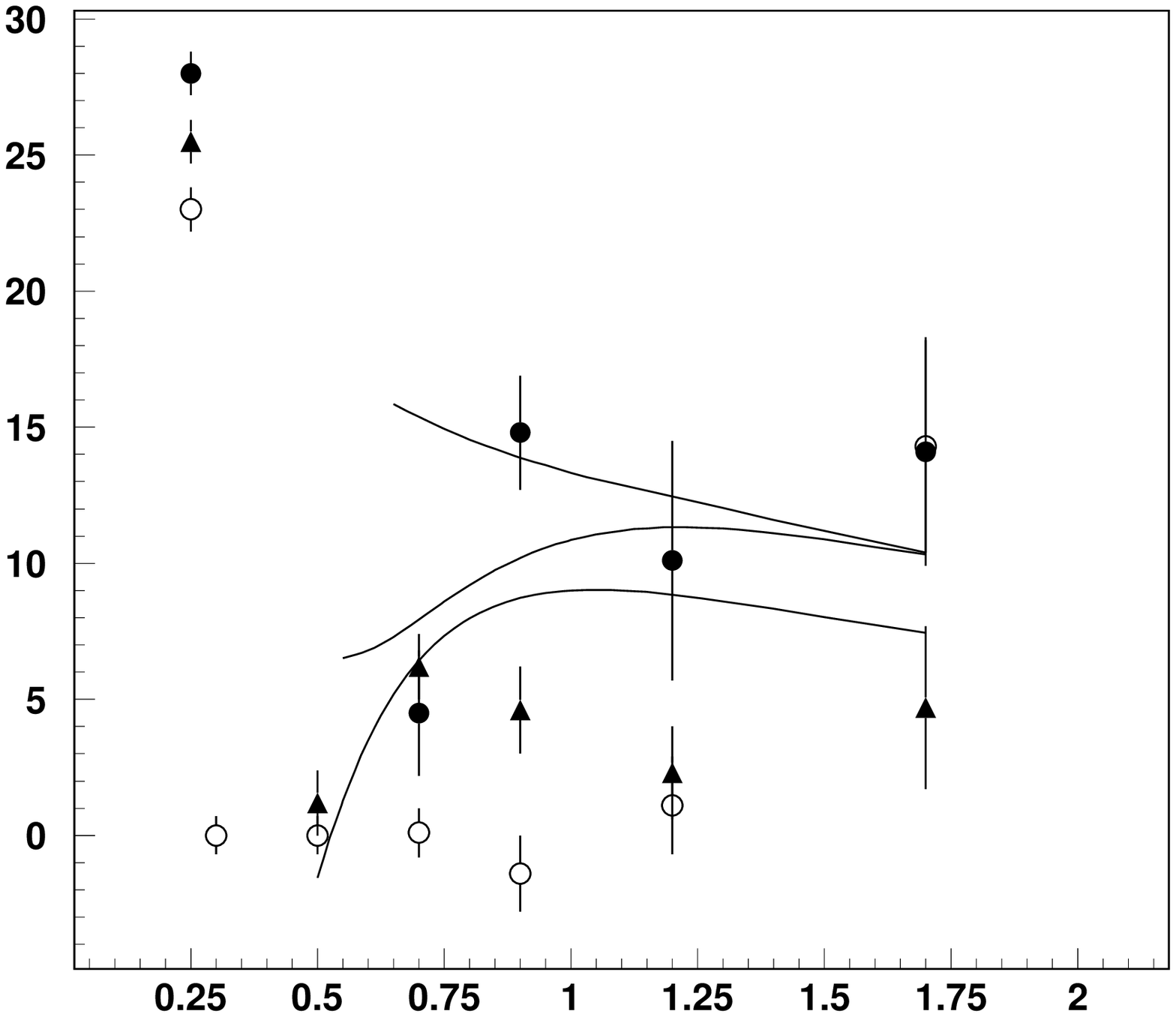}
  }
\end{picture}
    \caption{\label{fig:curves}
      The model calculations for the asymmetry plotted as a function
      of (a) $l_\bot$ and (b) $x$.}
  \end{center}
\end{figure}

The differential cross-section in the Born approximation is
\begin{equation}
  \frac{d\sigma_B}{dx\,dr}
  \;\, = \;\,
  \frac{2\alpha_{QED}\alpha_{pp\pi}}{m^2}
  \frac{r}{x(1+\rho)^2}
  \left[ \ln\frac{(1+\rho)^2}{\sigma}-1 \right],
\end{equation}
where $\alpha_{pp\pi}=g^2/(4\pi)\approx3$, $\rho=r^2/x^2$ and
$\sigma=(\mu^2+q_m^2)/M^2$.  The Born cross-section is a monotonically
rising function of $x$ (${\sim}x^3$) for small $x$.  It falls rapidly
as $1/l^3$ for large pion transverse momentum $\vec{l}$ and reaches
the maximum value for $l_\bot\sim{}mx/\sqrt{3}$.  The asymmetry in the
inclusive set-up, defined as a ratio of even and odd parts of the
cross-section averaged over transverse momenta, may be written, for
small $\sigma$, in form:
\begin{equation}
  A
  \;\, = \;\,
  \frac{\int d^2\vec{q} \, \sum\Delta|M|^2}
       {\int d^2\vec{q} \, \sum      |M|^2}
  \;\, = \;\,
  |\vec{a}| \sin\phi_l \,
  \frac{4\alpha_{\QED}R}
       {r\left[ \ln\frac{(1+\rho)^2}{\sigma}-1 \right]},
\end{equation}
with
\begin{eqnarray}
  R
  &=&
  x \ln(1+\rho) \ln\frac{1+\rho}{\sigma}
\nonumber\\
  && +x(1-x)
  \left[
  \frac{\rho}{2(1+\rho)} \ln^2\sigma
  +
  f_1(\rho,x) \ln\sigma
  +
  f_2(\rho,x)
  \right],
\label{eq:R}
\end{eqnarray}
where $f_{1,2}$ are rather flat functions; the complete expression for
$R$ is given in the Appendix.  The results of a numerical calculation
of the asymmetry in the inclusive set-up as a function of $\vec{l}$
for a various of values of the Feynman variable, $x$, and as a
function of $x$ for various values of $\vec{l}$ are presented in
Fig.~\ref{fig:curves}.  While the detailed dependence on $x$ and 
$l_\bot$ is not entirely reproduced, the general trends are seen to be
correct.

A similar calculation for the case of $\pi^+$ production in the
small-$\vec{q}$ limit yields
\begin{eqnarray}
  x^2     \sum       |\mathcal{M}_{\pi_+}|^2 \hphantom{\Delta}
  &=&
  (1-x)^2 \sum       |\mathcal{M}_{\pi_0}|^2,
\nonumber\\
  x^2     \sum \Delta|\mathcal{M}_{\pi_+}|^2
  &=&
  (1-x)^2 \sum \Delta|\mathcal{M}_{\pi_0}|^2,
\end{eqnarray}
where $\sum\Delta|\mathcal{M}|^2$ stands for the spin-weighted sum.
Thus, the asymmetries for $\pi^+$ production roughly coincide with
those for the $\pi^0$ case.

\section{Conclusions}

We see that asymmetry effects due to single transverse polarisation
are not suppressed in the limit of large total CMS energy, $\sqrt{s}$,
in the case when the produced hadron belongs to the jet of the
polarised proton.
The overall normalisation depends on the detailed mechanism of vector
meson (photon and gluon) interaction with nucleons and is bound to the
choice of the parameters $\alpha$ and $\mu$. The naive replacements
$\alpha\to\alpha_s$ and $\mu\to\Lambda_\QCD$ lead to asymmetries of
the same order as those found experimentally and which grow with
transverse momentum for small values. This behaviour agrees
qualitatively with the existing data \cite{Adams:1991cs, Adams:1992ru,
  Adams:1991rw, Adams:1997wf}.

It should be noted that in the case when the pion is detected in the
direction of the jet moving in the opposite direction (i.e., along
$\vec{p}_1$) the asymmetry effect will be suppressed in the
$s\to\infty$ limit. In fact, information on the transverse
polarisation of proton $p_2$ cannot be transmitted to jet components
developing from proton $p_1$ unless at least one ``sense'' component
of the virtual photon polarisation tensor is used, $g_\bot^{\mu\nu}$;
consequently, it will be suppressed by powers of $m^2/s$.\footnote{We
  are grateful to Lev N.~Lipatov for discussions on this point.}

In particular, for elastic proton-proton scattering where an
unpolarised scattered proton with momentum $p_1'$ is detected we
obtain:
\begin{equation}
  A_{p(p_1')}
  \;\, = \;\,
  \alpha
  \frac{5m \vec{q}^2 (\vec{a}\vprod\vec{q})_z}
       {2s^2}.
\end{equation}

Higher-order QCD effects may be taken into account by introducing a
factor $(s/s_0)^{a_\Odderon}$ into the odd part of the elastic
proton-proton scattering zero-angle amplitude, where $a_\Odderon$ is
the odderon intercept.  Thus, the asymmetry considered here,
associated with twice the imaginary part, acquires a factor
$a^2(s/s_0)^{a_\Odderon-a_\Pomeron}$, where $a_\Pomeron$ is the
Pomeron intercept. For small-angle scattering of electrons off
polarised protons, the odderon contribution manifests itself in higher
orders of PT due to conversion of photons into gluons through the
$\gamma\gamma\to{}gg$ and $\gamma{g}\to{}gg$ kernels, which may also
be investigated at DESY.

\section{Acknowledgments}
One of us (E.A.K.) is grateful to Pierre Gauron, Basarab Nicolescu and
Gregory P. Korchemsky for useful discussions, also to the Cariplo
Foundation for Scientific Research, for financial support, and to the
Department of Sciences, University of Insubria, Como, for hospitality,
while this work was performed.

\appendix
\section*{Appendix}

After some simple operations, the odd part of the cross-section
averaged over $\vec{q}$ may be put into the form:
\begin{eqnarray}
  \int \frac{d^2\vec{q}}{\pi} &\displaystyle\sum& \Delta|M|^2
  \;\, = \;\,
  \frac{2^{10}(\pi g)^2\alpha_\QED^3xs^2}{d_2^2} \,
  \frac{|\vec{a}| \sin\phi_l}{r}
\nonumber\\
  && \times
  \int \frac{d^2\vec{q}}{\pi} \,
  \frac{\ln(\vec{q}^2/\mu^2) \vec{q}\cdot\vec{l}}
       {(\vec{q}^2+\mu^2)^2 d_1 \tilde{s}} \,
  [ x^2 \vec{q}^2 \tilde{s}
  - 4(\vec{q}\cdot\vec{l})^2
  - 2xd_2\vec{q}\cdot\vec{l} ],
\label{eq:Rint}
\end{eqnarray}
where we have used another equivalent form of the numerator in 
Eq.~(\ref{eq:asym}):
\begin{equation}
  [ x\tilde{s} - 2\vec{q}\cdot\vec{l} ]
  [ x\vec{q}^2 + (1-x)2\vec{q}\cdot\vec{l} ]
  \;\, = \;\,
    x^2\vec{q}^2\tilde{s}
  - 4(\vec{q}\cdot\vec{l})^2
  - 2xd_2\vec{q}\cdot\vec{l}
\end{equation}
We write the integrals of the three terms in square brackets in 
Eq.~(\ref{eq:Rint}) as $J_{1,2,3}$ respectively. The quantity $R$ 
introduced in Eq.~(\ref{eq:R}) may then be re-expressed in terms of 
the $J_i$:
\begin{equation}
  R = \frac{\sum_{i=1}^3J_i}{x(1-x)}.
\end{equation}

For the first term we have
\begin{equation}
  J_1
  \;\, = \;\,
  x(1-x) \int\frac{d^2\vec{q}}{\pi} \, \vec{q}\cdot\vec{l} \,
  \frac{\ln(1/\sigma)+\ln(\vec{q}^2/m^2)}
       {\vec{q}^2(\vec{q}^2-2\vec{q}\cdot\vec{l}/x+(1+\rho)m^2)},
\end{equation}
where $\rho=\vec{l}^2/(m^2x^2)$.  To use the Feynman trick of
combining the denominators, we use the following representation for
the logarithm:
$$
  \left(\frac{m^2}{\vec{q}^2}\right)
  \ln\left(\frac{\vec{q}^2}{m^2}\right)
  =
   \left.
  -\frac{d}{dg} \left(\frac{\vec{q}^2}{m^2}\right)^{-g}
   \right|_{g=1}.
$$
The denominators may be combined using the identity
\begin{equation}
  \frac{1}{u^g v^h}
  \;\, = \;\,
  \frac{\Gamma(g+h)}{\Gamma(g)\Gamma(h)}
  \int_0^1 dz \, \frac{(1-z)^{g-1}z^{h-1}}{[(1-z)u+zv]^{g+h}}.
\end{equation}
The further standard procedure of performing the $d^2\vec{q}$
integration, subsequent differentiation with respect to $g$ and
integration over $z$ yields
\begin{equation}
  J_1
  \;\, = \;\,
  x^2(1-x) \ln(1+\rho) \ln\frac{1+\rho}{\sigma}.
\end{equation}

In the evaluation of $J_2$ we may set $\sigma=0$ in the denominator.
Joining the first two denominators we have
\begin{equation}
  \frac{1}{d_1\tilde{s}}
  \;\, = \;\,
  \frac{(1-x)^2}{x}\int_0^1 dy \,
  [\vec{q}^2-2\vec{q}\cdot\vec{l}\eta/x + \eta m^2(1+\rho)]^{-2},
\end{equation}
where $\eta=x+y(1-x)$.  And following a procedure similar to that
given above we obtain
\begin{eqnarray}
  J_2
  &=&
  -4[x(1-x)\rho]^2\int_0^1 dy \, \eta \int_0^1 dz \, z^2(1-z)
\nonumber\\
  &&\hspace*{5cm}\times
  \left[ \frac{\rho\eta^2 z^2}{D^3}(2L-1)+\frac{3L}{2D^2} \right],
\end{eqnarray}
where $L=\ln\frac{D}{(1-z)\sigma}$ and
$D=\eta(1+\rho)z-\eta^2z^2\rho+(1-z)\sigma$.  We note that we may set
$\sigma=0$ in the expression for $D$ although in evaluating $J_3$ we
cannot omit $\sigma$ in the denominator. Nevertheless, using the
identity
\begin{equation}
  \ln\left(\frac{\vec{q}^2}{\sigma m^2}\right)
  \;\, = \;\,
    \ln\left(\frac{\vec{q}^2/m^2+\sigma}{\sigma}\right)
  - \ln\left(1+\frac{\sigma m^2}{\vec{q}^2}\right),
\end{equation}
we may apply the procedure of differentiation to the first term. The
second is important in the region $\vec{q}^2\sim\sigma{m^2}$ and may
be evaluated explicitly:
\begin{eqnarray}
  &&
  -2x^2(1+\rho) \int\frac{d^2\vec{q}}{\pi} \, (\vec{q}\cdot\vec{l})^2
  \frac{\ln(1 + \frac{\sigma m^2}{\vec{q}^2})}
       {(\vec{q}^2/m^2 + \sigma)^2 d_1 \tilde{s}}
\nonumber\\
  && \hspace{3cm} \;\, = \;\,
  -\frac{x^2(1-x)^2\rho}{1+\rho}
  \int_0^\infty \frac{dz \, z \ln(1+\frac{1}{z})}
                     {(1+z)^2}
\nonumber\\
  && \hspace{3cm} \;\, = \;\,
  -\frac{x^2(1-x)^2\rho}{1+\rho}\left(\frac{\pi^2}{6}-1\right).
\end{eqnarray}
The total answer for $J_3$ is then
\begin{eqnarray}
  J_3
  &=&
  -\frac{x^2(1-x)^2\rho}{1+\rho} \left(\frac{\pi^2}{6}-1\right)
\nonumber\\
  &+&
  2x^3(1-x)^2 \rho(1+\rho) \\
  & &
  \times \int_0^1 dy \int_0^1 dz \, z(1-z)
  \left[ \frac{\rho z^2\eta^2(2L-1)}{D^3}
       + \frac{L}{2D^2} \right].
\qquad
\end{eqnarray}


\end{document}